\UseRawInputEncoding

\documentclass[letterpaper]{article} 
\usepackage{aaai25}
\usepackage{times}  
\usepackage{helvet}  
\usepackage{courier}  
\usepackage[hyphens]{url}  
\usepackage{graphicx} 
\usepackage{amsmath} 
\usepackage{adjustbox}
\usepackage{tcolorbox}

\usepackage{natbib}  
\usepackage{caption} 
\usepackage[
    colorinlistoftodos,
    textsize=footnotesize,
        ]{todonotes}

\usepackage[
    colorinlistoftodos,
    textsize=footnotesize,
        ]{todonotes}
        
\definecolor{myblue}{RGB}{10, 70, 180}  

\usepackage{xcolor}

\usepackage{graphicx}
\usepackage{booktabs}
\usepackage{pifont} 

\newcommand{\cmark}{\ding{51}}%
\usepackage{newfloat}
\usepackage{float} 
\usepackage{listings}
\DeclareCaptionStyle{ruled}{labelfont=normalfont,labelsep=colon,strut=off} 
\floatstyle{ruled}
\newfloat{listing}{tb}{lst}{}
\floatname{listing}{Listing}
\title{Contextualizing Recommendation Explanations with LLMs: \\ A User Study}

\author {
    Yuanjun Feng\textsuperscript{\rm 1},
    Stefan Feuerriegel\textsuperscript{\rm 2},
    Yash Raj Shrestha\textsuperscript{\rm 1}
}
\affiliations {
    \textsuperscript{\rm 1}University of Lausanne\\
    \textsuperscript{\rm 2}Munich Center for Machine Learning \& LMU Munich\\
    yuanjun.feng@unil.ch, 
    feuerriegel@lmu.de, 
    yashraj.shrestha@unil.ch
}

\begin{document}

\maketitle

\begin{abstract} 
Large language models (LLMs) are increasingly integrated into recommender systems to generate natural language explanations for contextualized content. This study examines how different types of LLM-generated explanations shape user perceptions and behaviors in the context of movie recommendations. In a pre-registered, between-subject online experiment (\emph{N}=759) and follow-up interviews (\emph{N}=30), we compare LLM-generated (a) generic explanations, and (b) contextualized explanations. We find that contextualized explanations better address users’ cognitive and affective needs, and increase their intention to watch the recommended movies. Insights from interviews reveal that contextualization fosters trust but depends heavily on alignment with users’ genre and narrative preferences. Notably, including more historical movie references improves user perceptions and intent, though overly detailed explanations can feel redundant. Users with higher engagement or trust in the system benefit most from contextualization. These findings highlight the potential of LLMs to enhance user-centered recommendation experiences. We offer practical guidance for designing explanation strategies that balance personalization and cognitive effort to improve relevance, trust, and engagement across social media and entertainment platforms.\end{abstract}

%
\begin{links}
    \link{Code}{https://anonymous.4open.science/r/Contextualizing-Recommendation-Explanations-With-LLMs-55D3}
\end{links}

\section{Introduction}

Recommender systems influence human decision-making in daily life, such as purchasing products on e-commerce platforms and choosing movies on streaming services \cite{Angelov.2021, Goyani.2020}. However, users often fail to understand why they receive specific recommendations. On streaming platforms like Netflix and Disney+, users are presented with recommended movies without any explanation of why those movies might be of interest. Without such explanations, users may find it difficult to judge whether the recommendation is helpful and decide on their consumption. Explanations help users make choices that reflect their preferences and improve decision quality \cite{Vultureanu.2021}. As a result, there is a growing research interest in offering explanations behind recommendations \cite{Ko.2022, Zhang.2020}. 

The emergence of large language models (LLMs) presents potential for improving recommendation explanations \cite{wu.SurveyLargeLanguage.2024, hou_chatgpt_2024}. Some studies have adopted LLMs to produce natural and human-like explanations. However, current efforts often rely on generic explanation templates \cite{LeiLi.Generate.2020}. These generic explanations overlook users' contexts or preferences, thereby limiting the benefits of the explanations and underutilizing the LLMs' reasoning capabilities. Existing research seldom provides a comprehensive understanding of how users perceive the received recommendations. Current work also rarely examines whether positive user perceptions translate into users' consumption intentions.

Motivated by this gap, our study adopts established explanation measurement standards \cite{Balog.Conflicting.2020} and interprets user responses through the lens of the \textbf{Uses and Gratifications (U\&G) framework} \cite{Katz.1973, lin_bigbirds_2013}. The U\&G framework categorizes users' diverse needs, including the desire for understanding and knowledge acquisition (\textbf{Cognitive needs}), the need for emotional connection and engagement (\textbf{Affective needs}), and the focus on practicality and task-oriented goals (\textbf{Utilitarian needs}). This framework provides a valuable perspective to understand how users actively select, consume, and engage with the received information. Our goal is to examine how different LLM-generated recommendation explanation types resonate with specific user needs and, in turn, influence their consumption intentions. Therefore, we designed the following research question:

\begin{tcolorbox}[colframe=black, colback=white, boxrule=0.2mm]
\textbf{How do LLM-generated explanations respond to users' diverse motivational needs (cognitive, affective, utilitarian) and affect their consumption intentions?}
\end{tcolorbox}

\begin{figure*}[htbp]
  \centering
  \includegraphics[width=1.0\textwidth]{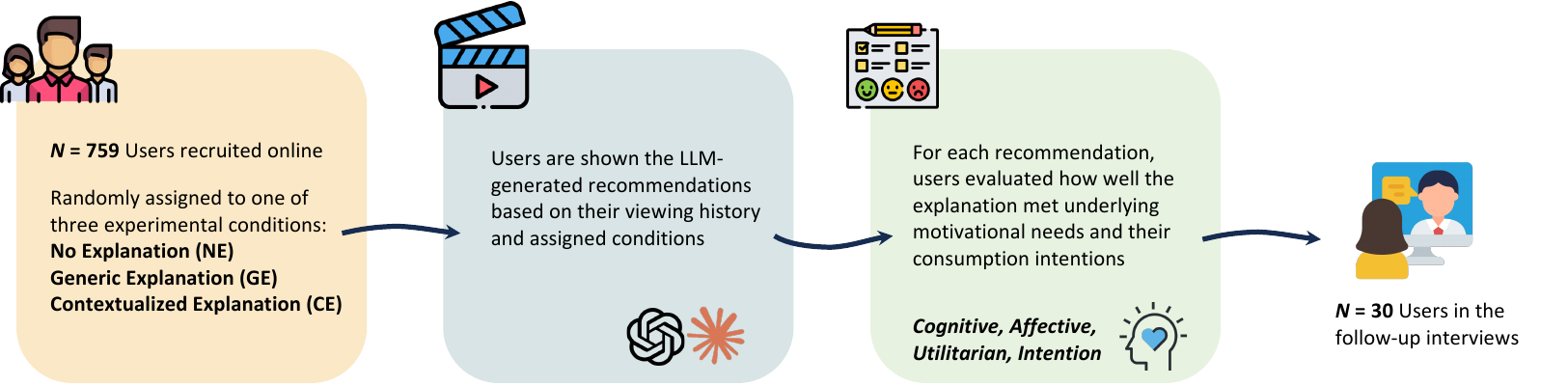}
  \caption{Overview of the study process.}
  \label{fig:study_flow}
\end{figure*}

In this study, we conducted a pre-registered, between-subject online experiment involving $N$=759 subjects from the U.S. in the context of movie recommendations.\footnote{The experimental protocol was pre-registered at OSF: https://osf.io/j46vx/?view\_only=dc2b1f641b33407ab8168736a783055a}
By drawing upon the U\&G framework, we analyze users' perceptions and intentions towards three types of recommendations generated by GPT-4 Turbo: (1)~recommendations with no explanations (\textbf{NE}), (2)~recommendations with LLM-generated generic explanations (\textbf{GE}), and (3)~recommendations with LLM-generated contextualized explanations (\textbf{CE}).\footnote{We focus on real-world recommendation tasks based on viewing histories without relying on explicit or private user preferences (e.g., as in cold-start problems of recommendation websites). As such, we use the term \emph{contextualized} rather than \emph{personalized} throughout the paper. This terminology aligns with related work \cite{bar_generative_2024} and reflects practical usage on platforms such as IMDb, Yelp, and Google Maps, where recommendations are often informed by user activity (e.g., viewing or browsing history) but not by direct feedback like ratings.} We also drew upon user profile information and conducted follow-up interviews ($N$=30) to interpret our results more deeply.
To ensure the robustness of our study design and the generalizability of the results, we conducted supplementary experiments by using a different LLM (Claude 3.5 Sonnet) with an additional 140 users from the U.S. Figure~\ref{fig:study_flow} provides an overview of our study flow.

\textbf{Contributions:} This paper demonstrates how LLM-generated, contextualized explanations can more effectively address users' diverse motivations and increase their consumption intentions. By integrating the U\&G framework, our study shows how different explanations affect users' cognitive, affective, and utilitarian needs. Moreover, we highlight the amplifying role of user engagement with the recommender system, revealing that users who already find the system beneficial gain the most from explanations. These findings bridge advancements in LLMs with user behavior and provide actionable insights to enhance user-centric recommender systems in social media contexts.

\section{Related Work}

\subsection{LLM-Generated Explanations in RecSys}
In social media and online communities, recommendations help users find items aligned with their preferences. Initial research centered on improving predictive accuracy \cite{Ma.ResearchDiversityAccuracy.2023, Wu.SurveyAccuracyOrientedNeural.2023}. However, growing attention to user trust and transparency has driven a shift toward explainable recommendations \cite{Burkart.SurveyExplainabilitySupervised.2021}. Initial research centered on improving accuracy \cite{Ma.ResearchDiversityAccuracy.2023, Wu.SurveyAccuracyOrientedNeural.2023}, but recent work emphasizes explainability to foster trust, transparency, and user satisfaction \cite{Burkart.SurveyExplainabilitySupervised.2021}. Traditional recommendation techniques (e.g., matrix factorization with interpretable layers \cite{Abdollahi.2017} or attention-based neural models \cite{Seo.2017, ChenChong.2018}) integrated explainability into model architectures.  

The advent of LLMs has made it easy to generate explanations in natural language more flexibly. Building upon template-based methods \cite{sarwar.ItembasedCollaborativeFiltering.2001, pazzani.ContentBasedRecommendationSystems.2007, yang.EffectsPopularityBasedNews.2016}, studies have used LLMs to generate detailed item descriptions \cite{Acharya.2023}, visual explanations \cite{Tsai.2019}, sentiment-oriented narratives \cite{chen_user_2019}, and interactive user-friendly justifications \cite{deldjoo_understanding_2024}. Previous works have also shown the unique potential of natural language explanations in mitigating information overload and facilitating community integration \cite{Chang.2016}. This growing trend of work underscores the potential for LLM-driven recommendation strategy in strengthening the recommendation processes.

\subsection{User Perceptions of Explanations}

Users increasingly rely on recommender systems to navigate overwhelming volumes of information. Grounded in the Uses \& Gratifications framework, individuals seek to satisfy their cognitive, affective, and utilitarian needs when engaging with social media content \cite{Katz.1973, Korhan.2016, bashardoust_comparing_2024}. Accordingly, explanations that clarify \emph{why} a recommendation is made can reduce cognitive effort, foster trust, and enhance decision confidence \cite{Mandl.ConsumerDecisionMaking.2011, Jameson.HumanDecisionMaking.2015}. 

Extant research also identifies several factors shaping user perceptions of explanations. For instance, users may prefer human-generated recommendations, particularly for hedonic products \cite{Wien.2021}. Explanations that reflect prior user feedback or behavior tend to resonate with individuals seeking tailored, value-reinforcing content \cite{LuHongyu.2023}. Moreover, individual differences in user profiles, such as cognitive engagement level, highlight the importance of detailed and transparent justifications for users who prioritize control and understanding \cite{Chatti.2022}. Therefore, aligning explanation strategies with varied user motivations and perceptions is essential for designing effective and user-centric recommender systems.


\subsection{Empirical User Studies on Explanations}

Empirical research has analyzed how different explanations in recommender systems impact users (see Table~\ref{tab:comparison_table}). These studies typically use controlled experiments and surveys to evaluate diverse explanation strategies. \citet{Silva.2024} conducted a user study (\emph{N}=94) comparing ChatGPT-generated generic and personalized movie explanations. While both were well-received, personalization improved perceived effectiveness when recommendations were less relevant and enhanced persuasiveness in low-confidence scenarios. Similarly, \citet{Lubos.2024} ran an online user study (\emph{N}=97), finding that LLM-generated explanations outperformed template-based baselines in trust and satisfaction due to their contextual richness and fluency. In another user study of a fitness plan recommender system (\emph{N}=341), \citet{sun_when_2023} showed that users trusted less of recommendations when explanations revealed reliance on their social-media friends’ activities, citing privacy and identity concerns. Instead, users preferred personalization based on their behavior. Together, these empirical findings provide insights into user perceptions and inform system design, but a comprehensive evaluation analyzing the cognitive, affective, and utilitarian needs is missing.

\subsection{Research Gap}

While the importance of recommendation explanations is widely acknowledged, research seldom examines how users interpret and respond to them. Furthermore, extant studies often rely on generic templates and have not fully explored the potential of LLM-generated contextualized explanations to address users' cognitive, affective, and utilitarian needs (see Table~\ref{tab:comparison_table}). To address this gap, our study provides the first comprehensive user evaluation of LLM-generated contextualized explanations through the lens of the U\&G framework. By examining how these explanations resonate with users' multifaceted needs and influence their intentions, we aim to offer deeper insights into designing recommendation explanations that are more aligned with user motivations.

\begin{table}[htbp]
\centering
\begin{adjustbox}{width=\columnwidth}
\begin{tabular}{lcccc}
\toprule
                          & Kunkel et al. & Silva et al. & Lubos et al. & \textbf{Our study} \\
\midrule
Contextualized explanations            &            & \cmark & \cmark & \cmark \\
Generic explanations                   & \cmark     & \cmark & \cmark & \cmark \\
\midrule
Cognitive needs           & \cmark     &        & \cmark & \cmark \\
Affective needs           & \cmark     & \cmark &        & \cmark \\
Utilitarian needs         &            & \cmark & \cmark & \cmark \\
\midrule
Consumption intentions        & \cmark     &        &        & \cmark \\
\bottomrule
\end{tabular}
\end{adjustbox}
\caption{Comparison of research focus between previous studies and our work.}
\label{tab:comparison_table}
\end{table}

\section{Method}

\subsection{Overview}
\label{sec:overview}
We conducted an online between-subjects experiment, where subjects ($N$=759) were randomly assigned to one of three experimental conditions: (1)~\emph{recommendations with no explanations} (\textbf{NE}), (2)~\emph{recommendations with LLM-generated generic explanations} (\textbf{GE}), and (3)~\emph{recommendations with LLM-generated contextualized explanations} (\textbf{CE}).

In the experiment, users first selected movies they had previously watched from a given set to establish the viewing history, based on which LLMs generated movie recommendations and explanations depending on the experimental condition.\footnote{We used GPT-4 Turbo (version: gpt-4-1106-preview), which has knowledge of world events up to April 2023. For the experiment in the supplement, we used Claude 3.5 Sonnet.} After receiving the corresponding explanation, users provided a series of ratings reflecting how well the explanation met underlying motivational needs, as well as their resulting behavioral intentions:

\begin{itemize} 
\item \textbf{Cognitive needs}: reflecting users' desire for understanding and credibility in the recommendation process, assessed through \emph{Scrutability}, \emph{Transparency}, and \emph{Trust}. 
\item \textbf{Affective needs}: addressing the emotional and motivational dimensions of the user experience, evaluated through \emph{Persuasiveness} and \emph{Satisfaction}. 
\item \textbf{Utilitarian needs}: focusing on practical, goal-oriented aspects of the recommendation, captured by \emph{Effectiveness}, \emph{Efficiency}, and \emph{Helpfulness}. 
\item \textbf{Consumption intentions}: reflect users' behavior outcome, captured by \emph{UserIntent}.
\end{itemize}
Dimensions composing the needs and intentions are established for assessing explanations in prior user studies \cite{Balog.Conflicting.2020}. We organize the dimensions under the U\&G framework to depict users' active pursuit of various needs from explanations. Given the statement for each dimension, users rated their agreement on a 7-point Likert scale from 1 (``Extremely Disagree'') to 7 (``Extremely Agree''). The higher-level intentions and needs are averaged on the specific dimensions (e.g., \emph{Cognitive} is the average of \emph{Scrutability}, \emph{Transparency}, and \emph{Trust}). A summary of the main variables and survey questions is in Table~\ref{tab:variablesummary}. After the survey, we randomly invited users to conduct follow-up interviews ($N=30$) to see their latent motivations across different explanations.

\begin{table}[htbp] 
\centering
\scriptsize 
\begin{tabular}{ll}
\toprule
\textbf{Variable} &  \textbf{Questions} \\
\midrule 
\multicolumn{2}{l}{\textsc{Cognitive Needs}}\\ 
\midrule
\emph{Transparency} & \multicolumn{1}{p{6cm}} {\emph{The text helps me to understand what the recommendation is based on.}} \\
\emph{Trust}  & \emph{The text helps me to trust the recommendation.} \\
\emph{Scrutability} & \multicolumn{1}{p{6cm}}{\emph{The text would allow me to give feedback on how well my preferences have been understood.}} \\

\midrule 
\multicolumn{2}{l}{\textsc{Affective Needs}}\\ 
\midrule
\emph{Persuasiveness} &  \emph{The text makes me want to watch this movie.} \\
\emph{Satisfaction} & \multicolumn{1}{p{6cm}}{\emph{The text would improve how easy it is to pick a recommendation.}} \\

\midrule 
\multicolumn{2}{l}{\textsc{Utilirian Needs}}\\ 
\midrule
\emph{Effectiveness} & \emph{The text helps me to determine how well I will like this movie.}  \\
\emph{Efficiency} & \emph{The text helps me to decide faster if I will like this movie.} \\
{\emph{Helpfulness}} & \emph{How helpful do you find the recommendation?} \\

\midrule 
\multicolumn{2}{l}{\textsc{Behvaiour Outcome}}\\ 
\midrule
{\emph{UserIntent}} & \emph{How willing are you to watch this recommended movie?} \\
\bottomrule

\end{tabular}
\caption{Summary of main variables and survey questions.}
\label{tab:variablesummary}
\end{table}

\subsection{Subjects}
\label{sec:subjects}

Subjects were recruited from the U.S. via Cint.\footnote{Cint: https://www.cint.com/} The online platform offers a built-in functionality for stratified sampling of subjects so that the sample is representative of the U.S. population. Subjects are remunerated with a fixed amount of around \$15 per hour.

As preregistered, we conducted a pilot study involving 50 subjects to assess the survey's functionality and design. Informed by other user studies \cite{Kang.SampleSizeDetermination.2021, lakens.SampleSizeJustification.2022}, we performed a power analysis using G*Power 3.1 software \cite{Faul.PowerFlexibleStatistical.2007, Faul.StatisticalPowerAnalyses.2009}. For multiple comparisons, we selected $F$-tests and chose a one-way {ANOVA} design with fixed effects \cite{lee.WhatProperWay.2018}. We set the significance level ($\alpha$) to 0.05 and the power (1-$\beta$) to 0.95, with the adjusted effect size set at 0.1. The power analysis based on the pilot study results suggested that we need at least 750 subjects for the experiment.

Initially, 1401 subjects completed the survey. We discarded responses from subjects failing an initial attention check. Subsequently, following prior research \cite{Pennycook.PriorExposureIncreases.2018} and our pre-registration, a subset of subjects were excluded for three reasons: (i)~subjects who self-reported to have answered randomly, (ii)~failed the second attention check, (iii)~responded excessively fast (i.e., the duration was less than around 6 seconds for responding to the user perceptions). These users still received remuneration after the exclusion. Eventually, due to our strict exclusion criteria, 759 of the 1,401 subjects qualified, exceeding our estimated sample size of 750 to ensure the results achieved the desired statistical power.

Among the 759 subjects, 387 are women, 370 are men, and two are non-binary. The sample has a mean age of 52.85 (SD=16.01) and a range of 70 years (18 to 88). Most subjects have at least a high school diploma or equivalent.

\subsection{Materials}
\label{sec:material}

We selected GPT-4 Turbo to generate movie recommendations for the main experiment. We conducted an additional experiment with Claude-3.5-Sonnet to validate the robustness of our study design and the generalizability of our findings (see Appendix~\ref{appendix:claude}). We pre-compiled all recommendations and explanations in a corpus to ensure no variability in the outcomes across different LLM calls. The corpus is generated based on the pre-defined strategy and prompts below. Therefore, \textbf{users with the same viewing history and assigned conditions received the identical recommendation explanation.}

\subsubsection{Recommendation Strategy}
We implemented a sequential recommendation strategy following the algorithm in Table~\ref{tab:corpus_generation} in Appendix~\ref{appendix:corpus_generation}. To reduce bias, users only rated movies they had not watched. For each user, we began the recommendation process with 10 distinct popular movies from the IMDb Top 250 list\footnote{https://www.imdb.com/chart/top/}. If the user had already watched a movie, we proceeded to new recommended movies until they reached an unwatched one. The recommended movies watched by users were also added to their viewing history and used to generate new recommendations. For example, if a user has watched \emph{Movie A} before, LLMs will recommend \emph{Movie B}. If the user has watched it as well, then LLMs will combine both movies in the explanation for \emph{Movie C}. This process continued until either an unwatched recommendation was found or a maximum of 4 recommendations were generated.

Rather than relying on explicit subjective feedback (e.g., likes or dislikes), we collected information about whether users had previously watched each movie. This design simulates a cold-start scenario \cite{lam_addressing_2008,chaimalas_bootstrapped_2023}, where prior viewing history is unavailable and where the system starts with general popular choices before adapting to observed behavior. This approach mirrors real-world conditions, where explicit user feedback is often sparse or unavailable. 

Recommender websites (such as IMDb, Rotten Tomatoes, Google Maps, Yelp, and TripAdvisor) are typically user-independent and free of personalization, either because they have no access to user-specific information or personalization is unwanted. They only have access to consumption data from their registered users, that is, actual review contributors. For Yelp, it is estimated that only 0.67\% of users are registered contributors, while the remaining users are visitors.\footnote{See \url{https://askwonder.com/research/percentage-yelp-users-mainly-write-reviews-restaurants-whrwmul7u}, accessed March 12, 2025.} Moreover, the main selling point of recommender websites like IMDb and Yelp is the comparison of products or services. As a result, users are not required to sign in to use the website. Forcing users to register on the website would probably be counterproductive and decrease activity, as users might not want to sign up due to privacy reasons. Consistent with these, we collect users' viewing histories---rather than explicit preferences---to generate recommendations and explanations.



\subsubsection{Prompt Generation}
We followed previous practices \cite{giray.PromptEngineeringChatGPT.2023} when designing the prompts. Table~\ref{tab:explanation_example} shows examples of contextualized and generic explanations. All explanations were limited to 100 words, aligning with the length of typical movie descriptions. To manage reading speed and cognitive load, each explanation was designed to include around three key features, such as the director, cast, plot, or scenes. Further, the length of the explanations was similar across the different conditions to ensure comparability.

The main difference between CE (contextualized explanations) and GE (generic explanations) is in the explicit connection: while both explanations incorporate various elements (e.g., feature, popularity, similarity) to support the recommendations, the contextualized explanations naturally integrate the recommended movies with the user's viewing history, explicitly highlighting how the user's inputs shape the recommendation. Meanwhile, the generic explanations focus exclusively on the recommended movies. All generated movies were cross-checked with the IMDB dataset\footnote{https://developer.imdb.com/non-commercial-datasets/} to ensure no hallucinations from the LLMs. Additionally, the corpus was validated by two experts with strong expertise in the field to ensure that all explanations used in the experiment were consistent and of fair quality.

\begin{tcolorbox}[colframe=black, colback=white, boxrule=0.2mm, title=System Prompt]
\small 
You are a movie recommendation assistant. Aim for clear, straightforward language that a broad audience can easily understand. 

Output each recommendation into three segments: The movie's details, formatted as ``Title - Release Year - Director"; A concise summary within 40 words that outlines the storyline; An explanation within 100 words.

The explanation should focus on concrete aspects such as specific character arcs, notable plot developments, or memorable scenes.
\end{tcolorbox}

\begin{tcolorbox}[colframe=black, colback=white, boxrule=0.2mm, title=User Prompt - CE]
\small
For \{Movie\}, provide one recommendation for a user with a viewing history: \{User viewing history\}. Highlight how it directly appeals to the provided viewing history.
\end{tcolorbox}

\begin{tcolorbox}[colframe=black, colback=white, boxrule=0.2mm, title=User Prompt - GE]
\small
For \{Movie\}, write a recommendation explanation.
\end{tcolorbox}

\begin{table}[htbp]
  \centering
  \includegraphics[width=0.48\textwidth]{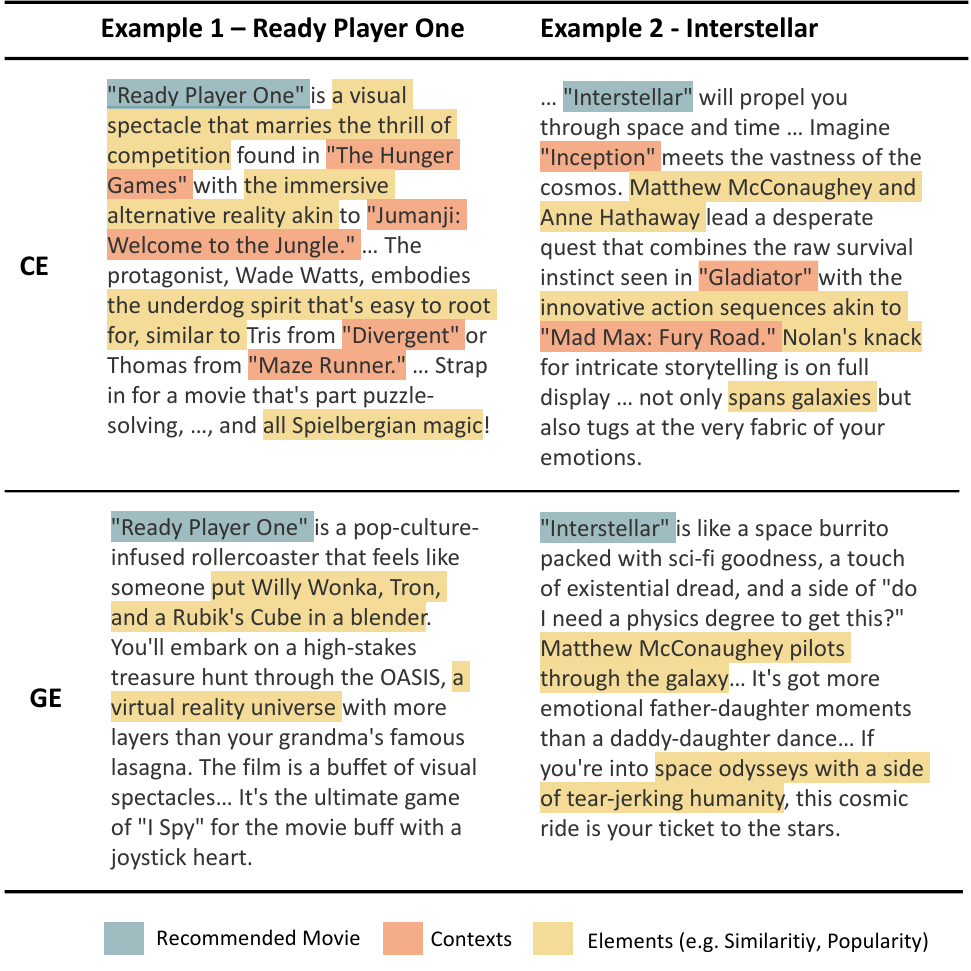}
  \caption{Examples of contextualized and generic explanations generated by GPT-4 Turbo.}
  \label{tab:explanation_example}
\end{table}

\subsection{Procedure}
\label{sec:procedure}

During the experiment, we collected users' viewing history and assigned them to one of our three conditions. Based on the condition assignment and viewing history, we generated the LLM-generated recommendations for them and collected their evaluations. The entire procedure was conducted online, consisting of the following steps:

\emph{Step 1.} Users visit a welcome page where they provide informed consent to proceed with the experiment.

\emph{Step 2.} Users read a short paragraph about the movie recommendation and respond to an attention check. Users who fail to pass are re-directed to the end of the survey. The attention check also serves to assess English proficiency.

\emph{Step 3.} Users read about the purpose and functionality of a recommender system.

\emph{Step 4.} Users select movies they have seen from the movie shortlist $\mathcal{M}$.

\emph{Step 5.} Users provide the frequency of their movie consumption, their usage frequency, and the perceived helpfulness of recommender systems.

\emph{Step 6.} Users are randomly assigned to one of our three conditions and shown corresponding recommendations. 

\emph{Step 7.} Following each recommendation, users express their perceptions across the dimensions of cognitive, affective, and utilitarian needs on a 7-point Likert scale. Users also state the recommendation's helpfulness and express their intentions to consume the movie. Steps 6 and 7 are repeated for all movies selected by users from $\mathcal{M}$.

\emph{Step 8.} Users go through a second set of attention checks.

\emph{Step 9.} Users provide demographic information, i.e., age, gender, and education.

\emph{Step 10.} Users participate in an honesty test where they are asked whether they answered all questions honestly, while being guaranteed that compensation would not be contingent on their answers.

\emph{Step 11.} Users receive a debrief of the experiment and are informed that LLMs power all movie recommendations.

Afterward, we randomly reached out to $N$=30 users for the follow-up interviews.\footnote{Detailed interview questions in Appendix~\ref{appendix:interview}.} We integrated their responses to help interpret the quantitative analysis results and investigate how explanations affect user motivational needs.

\subsection{Analyses}

We conducted analyses based on user perceptions in the survey. Each participant could provide up to 10 ratings, resulting in a total of 2716 ratings from 759 participants in the main experiment (904 in NE, 911 in GE, and 901 in CE). We used hypothesis testing to identify differences in perceptions and consumption intentions across recommendation types, followed by mixed-linear regression to examine how user characteristics explain variation in both variables. Additionally, we conducted follow-up interviews to further interpret the quantitative results. 

\textbf{Hypothesis testing:} We hypothesized that there are significant differences in the perceptions and consumption intentions between users receiving different recommendations. We conducted hypothesis testing on the four dependent variables of interest: \emph{Cognitive}, \emph{Affective}, \emph{Utilitarian}, and \emph{UserIntent} across the different conditions. Since the data did not meet the normality assumption \cite{lilliefors.KolmogorovSmirnovTestNormality.1967a}, we used a \emph{Mann–Whitney test} with \emph{Holm correction} to identify pairwise differences between conditions \cite{mchugh.MultipleComparisonAnalysis.2011}. 
Later, we use $\mu$ to report the mean value.
 
\textbf{Mixed-effects linear regression.} We used linear mixed-effects regression models to check further if any factors explained the variations in the three variables of interest.\footnote{We also calculated the variance inflation factor (VIF) for all independent variables. All VIF values were below 5, indicating that multicollinearity should not bias the estimates \cite{akinwande.VarianceInflationFactor.2015}.} We set NE as the reference condition so that our \emph{Treatment} variables represent CE or GE. Given that each subject provides multiple ratings, we further included subject-specific random effects to control for between-subject heterogeneity \cite{linden.HeterogeneityResearchResults.2021}. Additionally, to control for the habituation effect caused by repetitive measurements \cite{gaito.RepeatedMeasurementsDesigns.1961, jankowski.HabituationEffectSocial.2021}, we included a control variable for the recommendation display order (i.e., \emph{DisplayOrder}). Our final regression model is:

\begin{small}
\begin{align}
Y_{ij} &= \beta_0  
+ \beta_1 \emph{Treatment}_{j} 
+ \beta_2 \emph{Age}_{j} 
+ \beta_3 \emph{Gender}_{j} 
+ \beta_4 \emph{Education}_{j}\notag\\
&\quad + \beta_5 \emph{MovieConsumption}_{j} 
+ \beta_6 \emph{RecSysFrequency}_{j} \notag\\
&\quad + \beta_7 \emph{RecSysUtility}_{j} + \beta_8 \emph{HistoryNum}_{j} + \beta_9 \emph{DisplayOrder}_{j} \notag\\
&\quad + \beta_{10} \emph{HistoryNum}_{j} \times \emph{MovieConsumption}_{j}\\
&\quad + \beta_{11} \emph{HistoryNum}_{j} \times \emph{RecSysUtility}_{j} 
+ u_{0j} + \epsilon_{ij} \notag
\end{align}
\end{small}

\noindent
where ${Y}_{ij}$ denotes the dependent variables (i.e., \emph{Cognitive}, \emph{Affective}, \emph{Utilitarian}, and \emph{UserIntent}) for the $i$-th observation of the $j$-th subject, with intercepts $\beta_0$, subject-level random effects $u_{0j}$, coefficients $\beta_1$ through $\beta_{11}$, and an error term $\epsilon_{ij}$.

\subsection{Ethical Considerations}
We respect the privacy and agency of all subjects potentially impacted by this work and take steps to protect their privacy. The experiment design has obtained approval from the Ethics Commission at the University of (name anonymized for review). All data were collected anonymously, and any personally identifiable features were removed. All steps during data collection and analyses followed standards for ethical research \cite{rivers.EthicalResearchStandards.2014}. We also informed users that movie recommendations and explanations were generated by an LLM. 

\section{Results}

Below, we state the results of the hypothesis tests. The regression results are also shown in Table~\ref{table:mixed_effect}.

\begin{figure*}[htbp]
  \centering
  \includegraphics[width=1.0\textwidth]{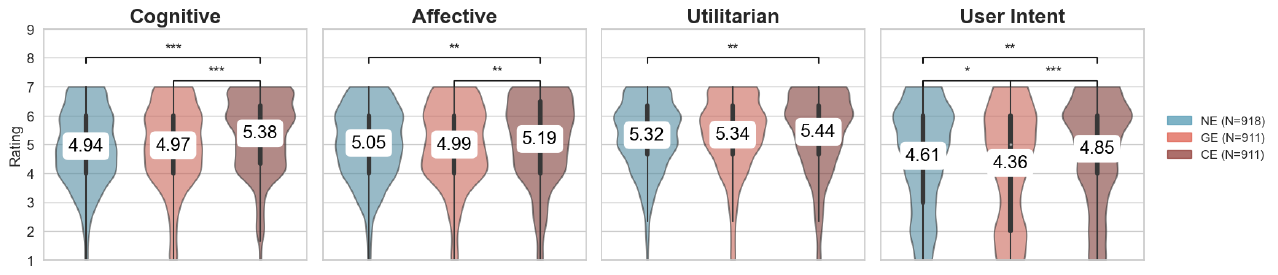}
  \caption{Ratings by explanation type (shown variable: \emph{Cognitive}, \emph{Affective}, \emph{Utilitarian}, and \emph{UserIntent}). Significance levels between conditions are: $^*p<0.05$, $^{**}p<0.01$, $^{***}p<0.001$.}
  \label{fig:violin}
\end{figure*}

\subsection{Cognitive Needs}
\label{sec:cognitive}

We observed a significant difference (Holm-adjusted $p<0.05$) in the distribution of the aggregated \emph{Cognitive} measure between NE ($\mu$=4.94), GE ($\mu$=4.97), and CE ($\mu$=5.38). Pairwise tests showed that CE significantly outperformed both NE and GE (Holm-adjusted $p$\textless0.05), indicating that contextual explanations better support users’ cognitive understanding. However, there was no statistical difference between NE and GE (Holm-adjusted $p$\textgreater0.05), suggesting that generic explanations offer no additional cognitive benefit over providing no explanation.

In the mixed-effects regression, both \emph{RecSysUtility} ($\beta$=0.304, $p$\textless0.05) and \emph{MovieConsumption} ($\beta$=0.150, $p$\textless0.05) demonstrated positive, significant effects on \emph{Cognitive}, indicating that participants who perceive the recommender system as more useful and who watch movies more frequently tend to rate higher in \emph{Cognitive}. Besides, more movie contexts ($\beta$=0.048, $p$\textless0.05) in the explanation also better satisfy users' cognitive needs.

In the interviews, several participants clearly expressed their trust and commended the clarity when receiving contextualized explanations.

\emph{``The thing that makes me trust the explanation was reading how the description had brought up other titles of movies that I saw in the recommendations, that had true things about the characters, and how the movies could be similar or how I could watch them already.'' (Female, 29)}

\emph{``They made the details of the movies sound more exciting and interesting and gave you more of a full rundown of theories so you don't think it's going to be one thing and turn out another.'' (Female, 31)}

On the other hand, sometimes users disliked the recommendations, but they still helped to filter out unwanted content, improving their sense of control and overall transparency. 

\emph{``The movies it recommended didn't interest me, so I don't think I'll be watching them. Felt like I understood the movies weren't for me.'' (Male, 32)}

\emph{``I trusted the recommended movie, but I made my own mind up after reading the explanation.'' (Male, 44)}



\subsection{Affective Needs}
\label{sec:affective}

For the \emph{Affective} dimension, CE ($\mu$=5.19) was rated significantly higher than both NE ($\mu$=5.05) and GE ($\mu$=4.99) (Holm-adjusted $p$\textless0.05). Still, no significant difference emerged between NE and GE (Holm-adjusted $p$\textgreater0.05).

The mixed-effects regression highlights two user-engagement variables, \emph{RecSysUtility} ($\beta$=0.295, $p$\textless0.05) and \emph{MovieConsumption} ($\beta$=0.175, $p$\textless0.05), as positive, significant predictors of affective responses. This suggests that, regardless of the explanation style, users who view the recommender system as helpful and frequently watch movies are more likely to have positive feedback in affective needs. Besides, more movie context ($\beta$=0.069, $p$\textless0.05) in the explanation has a positive effect as well.

From the interview, we observed that the movie and explanation quality are important to users' affection, no matter whether it contains the context. This also explains why users' satisfaction is fairly similar when receiving CE or GE. Regarding the textual quality, the output of LLMs is consistent regardless of the explanation types.

\emph{``Because I am a movie buff, and frankly, the descriptions were very well laid out so that it gave me enough information about the plot but also left enough unsaid that now I have to find out." (Male, 39)}

\emph{``Some of the word descriptions were a little over the top. The description was a little graphic." (Female, 68)} 

\emph{``The explanation was well written, so it gave away a little more than I would normally like, but was just as good as a trailer in my mind." (Male, 39)}



\begin{figure}[htbp]
  \centering
  \includegraphics[width=0.45\textwidth]{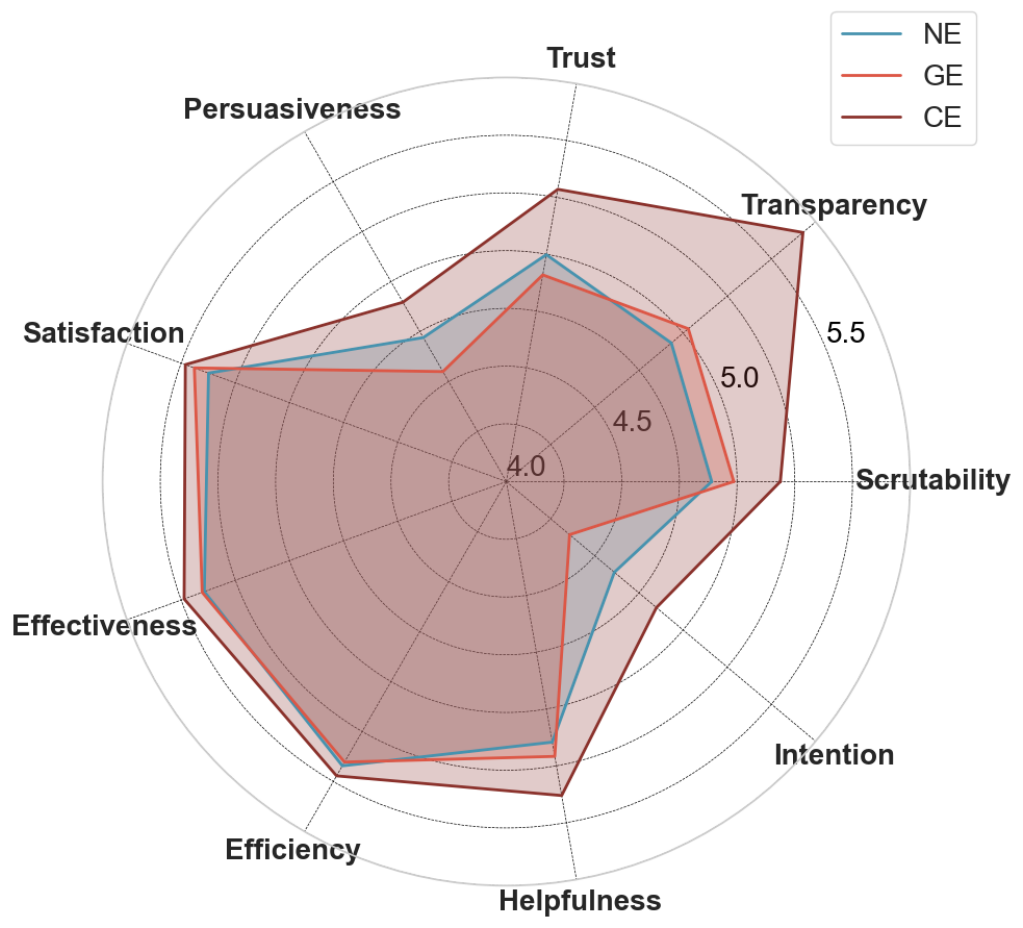}
  \caption{Radar chart of average user perception ratings in different dimensions by explanation types.}
  \label{fig:radar}
\end{figure}

\subsection{Utilitarian Needs}
\label{sec:utilitarian}

In the \emph{Utilitarian} dimension, CE ($\mu$=5.44) was rated significantly higher than NE ($\mu$=5.32) (Holm-adjusted $p$\textless0.05). There was no significant difference between NE and GE or CE and GE ($\mu$=5.34).

The mixed-effects regression showed \emph{RecSysUtility} ($\beta$=0.271, $p$\textless0.05) and \emph{MovieConsumption} ($\beta$=0.172, $p$\textless0.05) both positively influenced \emph{Utilitarian}, implying that users who closely engage with recommender systems and movies tend to see greater practical value in the recommendations regardless of explanation style. Female participants also reported higher utilitarian scores (versus male, $\beta$=0.292, $p$\textless0.05). More movie context increases utilitarian value ($\beta$=0.044, $p$\textless0.05).

In general, users felt that these explanations save time in deciding whether a movie fits their tastes. However, we observed that some users appreciated more details to support their decisions, while some preferred the explanations to be simplified.

\emph{``If I couldn't clearly understand the explanation of the film, I thought that I was most likely not going to enjoy the movie. I trust my first instincts on things because it has saved me a lot of time in the past, by allowing me to avoid seeing a movie that would only waste time in my too short life.'' (Female, 37)}

\emph{``Sometimes too much is overwhelming.'' (Male, 37)}

\emph{``The more detailed movie, the more it helped me decide.''(87, Female)}



\subsection{Consumption Intentions}
\label{sec:intention}

We found significant differences in the consumption intention across all conditions. CE ($\mu$=4.85) outperformed GE ($\mu$=4.36) and NE ($\mu$=4.61) (Holm-adjusted $p$\textless0.05). GE was also rated significantly lower than NE (Holm-adjusted $p$\textless0.05).

Mixed-effects regression indicates that \emph{RecSysUtility} ($\beta$=0.292, $p$\textless0.05) and \emph{MovieConsumption} ($\beta$=0.125, $p$\textless0.05) both increase the likelihood of watching the recommended title. Older participants ($\beta$=-0.147, $p$\textless0.05) expressed lower consumption intentions. Besides, more movie context ($\beta$=0.101, $p$\textless0.05) in the explanation has a positive effect as well.

Though users have overall positive feelings, their consumption intent is more attached to their taste. Many interviewees explicitly expressed that, if the movie is not their type, they will not watch it. This is not a consequence of our findings but of the accuracy of the algorithm behind the recommender systems. Hence, this still confirms the value of contextualization in motivating users' decisions.

\emph{``The factor that motivated me the most in deciding whether or not to watch the movie was the correlation factor between the new movie and movies I have watched in the past.''(Female, 20)}

\emph{``Sometimes the explanation was so intriguing I felt as though I wanted to know more about the movie. If the explanation is so boring, I don't finish it, I'm most likely not going to watch the movie.'' (Female, 37)}

\emph{``If it seemed like the movie was for children, it was a no for me.''(Male, 32)}

\emph{``This is becoming a little redundant, but again it explained it to the point I could see in my head, the picture and the action happening. Plus, anything with Will Smith has got to be awesome.'' (Female, 61)}



\begin{table}[ht]
\centering
\scriptsize
\begin{tabular}{p{2cm}cccc}
\toprule
 & \emph{Cognitive} & \emph{Affective} & \emph{Utilitarian} & \emph{UserIntent} \\
\midrule
\emph{Intercept} &
\begin{tabular}[c]{@{}c@{}}4.959***\\ (0.088)\end{tabular} &
\begin{tabular}[c]{@{}c@{}}5.030***\\ (0.087)\end{tabular} &
\begin{tabular}[c]{@{}c@{}}5.172***\\ (0.084)\end{tabular} &
\begin{tabular}[c]{@{}c@{}}4.849***\\ (0.112)\end{tabular} \\

\emph{Treatment (GE)} &
\begin{tabular}[c]{@{}c@{}}0.010\\ (0.102)\end{tabular} &
\begin{tabular}[c]{@{}c@{}}$-$0.032\\ (0.100)\end{tabular} &
\begin{tabular}[c]{@{}c@{}}0.064\\ (0.097)\end{tabular} &
\begin{tabular}[c]{@{}c@{}}$-$0.232\\ (0.127)\end{tabular} \\

\emph{Treatment (CE)} &
\begin{tabular}[c]{@{}c@{}}0.418***\\ (0.102)\end{tabular} &
\begin{tabular}[c]{@{}c@{}}0.166\\ (0.100)\end{tabular} &
\begin{tabular}[c]{@{}c@{}}0.165\\ (0.097)\end{tabular} &
\begin{tabular}[c]{@{}c@{}}0.277*\\ (0.127)\end{tabular} \\

\emph{Gender (Female)} &
\begin{tabular}[c]{@{}c@{}}0.120\\ (0.088)\end{tabular} &
\begin{tabular}[c]{@{}c@{}}0.160\\ (0.086)\end{tabular} &
\begin{tabular}[c]{@{}c@{}}0.292***\\ (0.084)\end{tabular} &
\begin{tabular}[c]{@{}c@{}}$-$0.181\\ (0.110)\end{tabular} \\

\emph{Age} &
\begin{tabular}[c]{@{}c@{}}$-$0.075\\ (0.046)\end{tabular} &
\begin{tabular}[c]{@{}c@{}}$-$0.089*\\ (0.045)\end{tabular} &
\begin{tabular}[c]{@{}c@{}}0.013\\ (0.044)\end{tabular} &
\begin{tabular}[c]{@{}c@{}}$-$0.147*\\ (0.057)\end{tabular} \\

\emph{Education} &
\begin{tabular}[c]{@{}c@{}}0.038\\ (0.042)\end{tabular} &
\begin{tabular}[c]{@{}c@{}}0.028\\ (0.041)\end{tabular} &
\begin{tabular}[c]{@{}c@{}}0.003\\ (0.040)\end{tabular} &
\begin{tabular}[c]{@{}c@{}}0.040\\ (0.052)\end{tabular} \\

\emph{RecSysUtility} &
\begin{tabular}[c]{@{}c@{}}0.304***\\ (0.063)\end{tabular} &
\begin{tabular}[c]{@{}c@{}}0.295***\\ (0.062)\end{tabular} &
\begin{tabular}[c]{@{}c@{}}0.271***\\ (0.060)\end{tabular} &
\begin{tabular}[c]{@{}c@{}}0.292***\\ (0.079)\end{tabular} \\

\emph{RecSysFrequency} &
\begin{tabular}[c]{@{}c@{}}0.019\\ (0.072)\end{tabular} &
\begin{tabular}[c]{@{}c@{}}$-$0.004\\ (0.070)\end{tabular} &
\begin{tabular}[c]{@{}c@{}}$-$0.023\\ (0.068)\end{tabular} &
\begin{tabular}[c]{@{}c@{}}0.132\\ (0.089)\end{tabular} \\

\emph{MovieConsumption} &
\begin{tabular}[c]{@{}c@{}}0.150**\\ (0.047)\end{tabular} &
\begin{tabular}[c]{@{}c@{}}0.175***\\ (0.046)\end{tabular} &
\begin{tabular}[c]{@{}c@{}}0.172***\\ (0.045)\end{tabular} &
\begin{tabular}[c]{@{}c@{}}0.125*\\ (0.059)\end{tabular} \\

\emph{DisplayOrder} &
\begin{tabular}[c]{@{}c@{}}0.040\\ (0.024)\end{tabular} &
\begin{tabular}[c]{@{}c@{}}0.035\\ (0.028)\end{tabular} &
\begin{tabular}[c]{@{}c@{}}0.017\\ (0.025)\end{tabular} &
\begin{tabular}[c]{@{}c@{}}0.115**\\ (0.042)\end{tabular} \\

\emph{HistoryNum} &
\begin{tabular}[c]{@{}c@{}}0.048**\\ (0.015)\end{tabular} &
\begin{tabular}[c]{@{}c@{}}0.069***\\ (0.018)\end{tabular} &
\begin{tabular}[c]{@{}c@{}}0.044**\\ (0.016)\end{tabular} &
\begin{tabular}[c]{@{}c@{}}0.101***\\ (0.028)\end{tabular} \\

\shortstack[l]{\emph{HistoryNum} \\ $\times$ \emph{MovieConsumption}} &
\begin{tabular}[c]{@{}c@{}}$-$0.028\\ (0.016)\end{tabular} &
\begin{tabular}[c]{@{}c@{}}0.007\\ (0.018)\end{tabular} &
\begin{tabular}[c]{@{}c@{}}$-$0.032*\\ (0.016)\end{tabular} &
\begin{tabular}[c]{@{}c@{}}$-$0.047\\ (0.029)\end{tabular} \\

\shortstack[l]{\emph{HistoryNum} \\ $\times$ \emph{RecSysUtility}} &
\begin{tabular}[c]{@{}c@{}}$-$0.019\\ (0.015)\end{tabular} &
\begin{tabular}[c]{@{}c@{}}$-$0.036*\\ (0.018)\end{tabular} &
\begin{tabular}[c]{@{}c@{}}$-$0.013\\ (0.016)\end{tabular} &
\begin{tabular}[c]{@{}c@{}}$-$0.054\\ (0.028)\end{tabular} \\
\bottomrule
\end{tabular}
\caption{Mixed-effects linear regression results. Coefficients are reported with standard errors in parentheses. Significance levels: $^*p<0.05$, $^{**}p<0.01$, $^{***}p<0.001$.}
\label{table:mixed_effect}
\end{table}

\subsection{Interpretations}

Based on the results above, we observed an overall advantage of LLM-generated contextualized explanations in addressing users’ motivational needs and increasing their consumption intentions. Figures~\ref{fig:violin} and~\ref{fig:radar} show that contextualized explanations successfully address users’ diverse needs. However, a key limitation remains: the explanations do not always align with users’ literary and genre preferences. Although we asked participants to evaluate the explanations rather than the recommended movies, interviews revealed that many users still evaluated the two together, resulting in lower scores for persuasiveness and consumption intention. When users turn to recommender systems, they often have specific goals, such as finding a movie to watch or exploring new options. Once the system delivers relevant recommendations, users may feel that their immediate needs are already satisfied. This likely explains why participants rated effectiveness, efficiency, and satisfaction high, even in the absence of explanations. Together, our findings highlight the value of LLM-generated explanations in building transparency and trust. However, to truly enhance user experience, explanations need to resonate with users’ tastes and expectations, especially in subjective domains like entertainment.

\textbf{Demographics and prior engagement shape user perception:} Based on the results in Table~\ref{table:mixed_effect}, users’ prior engagement significantly shapes their evaluation of explanations. Participants who consumed more movies rated explanations more positively across all dimensions, suggesting that domain familiarity enhances appreciation of contextualized content. Similarly, higher perceived usefulness of recommender systems strongly gave favorable evaluations.

Demographic effects were more selective. Female users gave higher utilitarian ratings, while older users rated affective and intention outcomes slightly lower. These findings suggest that both system engagement and user characteristics should be considered when designing a user-friendly system.

\textbf{More viewing history contexts, better perceptions:}
Table~\ref{table:mixed_effect} reveals that the number of viewing histroy (\emph{HistoryNum}) positively influenced users’ evaluations in perception and intention. As detailed in Section~\ref{sec:material}, we limited the number of contexts maximum of four, which was guided by typical movie description lengths, native reading speeds, and cognitive load thresholds.

Importantly, the non-significant interaction effects between \emph{HistoryNum} and both \emph{MovieConsumption} and \emph{RecSysUtility} suggest that the positive impact of including more movie contexts in explanations holds consistently, regardless of users' movie consumption frequency or perceptions of the recommender system’s helpfulness. This implies that richer contextualization can enhance user evaluations without being contingent on prior engagement. Although richer context tends to improve perception and intent, the effects, based on the coefficients in the regression results, remain relatively small. Some users also flagged certain explanations as redundant or overly detailed. To prevent cognitive overload, we should carefully calibrate how much context to include, ensuring that explanations remain concise yet informative.

\section{Heterogeneity Analysis}
\subsection{Claude 3.5 Sonnet}
To supplement the main experiment, we expanded the experiment to a larger scope by leveraging another LLM (Claude 3.5 Sonnet). Given that GPT-4 Turbo is one of the most powerful LLMs, we tested whether the superior performance of contextualization could be generalized to other LLMs. We generated another recommendation corpus using Claude 3.5 Sonnet and replicated the experiments on the U.S. samples ($N$=140, 71 are women, 68 are men, and one is non-binary). The results are aligned with our main findings, validating the advantages of LLMs in generating contextualized explanations in terms of meeting users' motivational needs and increasing their consumption intent. The results and interpretations are provided in Appendix~\ref{appendix:claude}. 

\subsection{Movie Genre}
To further understand how user perceptions vary across recommendation contexts, we examined the role of movie genre, specifically \textit{comedy} and \textit{adventure}, with supplementary mixed-effects regression models (see Equation~\ref{eq:mixed_effect_mv_genre} in Appendix~\ref{appendix:mv_genre}). As shown in Table~\ref{table:mixed_effect_genre}, comedy movies were consistently rated lower than adventure movies across affective, utilitarian, and consumption intention outcomes. We also observed that these genre effects were not moderated by explanation types. These supplementary results highlight genre as a key factor influencing user responses to LLM-generated recommendations. The influence of movie genre remains an important area for further investigation.
\vspace{1em}

In summary, the supplementary experiment yielded results that supported the main experiment's findings and conclusions, thereby validating the robustness of our study design and the generalizability of the findings.

\section{Discussion}

\subsection{Implications of Findings} Recommender systems play an important role in shaping user decisions, particularly influential in the context of web and social media. Given the far-reaching implications of these systems, there is growing demand from users and regulators for transparency, emotional connection, and efficiency in decision-making. This highlights the value of high-quality and context-aware recommendation explanations. However, many existing recommender systems do not or fail to provide explanations that address users' diverse needs \cite{LeiLi.Generate.2020, LuHongyu.2023}.

Our study demonstrates that LLM-generated contextualized explanations can overcome these limitations by satisfying users' cognitive needs, ultimately increasing their consumption intentions. Also, our findings reveal the insufficiency of LLM-generated explanations in meeting users' utilitarian and affective needs, which raises concerns about the proper design and implications of LLM-generated contextualized content.

\textbf{Build trust, but risk misrepresenting preferences:} Our results show that contextualized explanations significantly address users’ cognitive needs, particularly in terms of \emph{Transparency}, \emph{Trust}, and \emph{Scrutability}. By referencing users’ prior viewing history, LLM-generated explanations appear more tailored and meaningful, encouraging deeper engagement with recommendations.

Nevertheless, we also observed that LLMs tend to interpret viewing histories as an implicit preference signal. While this can increase perceived personalization, it also introduces a risk of over-interpretation, which is reflected in user interviews. In practice, real-world recommender systems often infer preferences from behavioral signals (e.g., clicks, watch time) rather than explicit feedback, which may not always align with users’ true interests or sentiments. LLMs, when generating explanations based on such data, may reinforce these weak or ambiguous signals, thereby creating a disconnect between the explanation and the user’s actual intent. This misalignment risks undermining the trust that contextualized explanations build. When users feel the system attributes preferences they do not hold, trust may erode over time. Therefore, in real-world deployments, systems that leverage LLMs for explanation generation should account for this interpretive bias and consider more nuanced strategies for referencing user history, particularly when preferences are inferred rather than explicitly stated.

\textbf{Opportunities and challenges for the movie industry and users:}
Our findings show that LLM-generated movie recommendations significantly increase users' intention to engage with suggested content. Effective recommendations can be generated by leveraging typical user viewing histories and state-of-the-art LLMs---without requiring additional user input or behavioral interventions. Such ease of implementation, coupled with LLMs' ability to produce highly contextualized content, makes them valuable tools for streaming services and other media platforms.


Integrating LLMs into recommendation systems offers the potential to enhance operational efficiency but introduces significant costs that vary by deployment strategy. Using commercial APIs such as OpenAI’s GPT-4o mini, priced at \$0.3 per million input tokens and \$1.2 per million output tokens\footnote{OpenAI API Pricing: \url{https://openai.com/api/pricing/}}, a typical user interaction (200 input and 100 output tokens) costs approximately \$0.00018. If each of Netflix’s 301.6 million global subscribers received 50 recommendations daily, the resulting API cost would amount to approximately \$2.7 million per day. These costs must be weighed against potential revenue gains. In subscription-based models like Netflix, the average revenue per user in North America was about \$17.17 in 2024 \cite{singh_netflix_2025}. In advertisement-driven platforms like YouTube, creators typically earn \$5 to \$7 per 1,000 ad views \cite{marshall_youtube_2024}. Even modest improvements in user engagement could yield substantial returns, emphasizing the importance of careful cost-benefit analysis when scaling LLM integration.

Further, while LLMs present substantial opportunities to enhance recommendation quality and operational efficiency, successful deployment requires careful management of associated technical, economic, and ethical challenges. In particular, LLM-generated recommendations carry risks related to biased outputs \cite{feuerriegel_generative_2023, bashardoust_effect_2024}. In movie recommendation contexts, biases may manifest as an underrepresentation of specific genres, cultures, or demographic perspectives. For example, the corpus generated in our study tended to favor Western-centric narratives disproportionately. Additionally, perceived biases or cultural insensitivities arising from skewed training data or adversarial prompt manipulation can erode user trust and pose significant reputational risks to platforms.

\subsection{Limitations}
While our study offers valuable insights into user perceptions of LLM-generated explanations for movie recommendations, several limitations should be noted.

First, our findings may be limited in generalizability due to both the participant sample and the underlying models. We primarily recruited English-speaking users, which may not capture the full range of cultural and linguistic differences in user perceptions. Additionally, the effectiveness of the explanations likely depends on the capabilities of the underlying LLMs. Although we used state-of-the-art models (GPT-4 Turbo and Claude 3.5 Sonnet), future advancements may further influence how users interpret and respond to recommendations and explanations.

Second, our study may oversimplify the explanation generation process. Future research should explore various explanation strategies to better understand their effectiveness across different contexts. While we employed a typical cold-start recommendation scenario, providing additional user information, such as demographics or expressed preferences, can more efficiently reveal LLMs’ strong contextualization capabilities.

Lastly, our experimental design does not account for the longitudinal dynamics of real-world user interactions with recommender systems. While we capture users’ immediate perceptions of explanations and recommendations, these impressions may evolve as users continue to engage with the recommended content over time.
Additionally, due to our control over user inputs, participants did not have the opportunity to express specific or niche movie preferences to the LLMs, which may play a significant role in shaping their perceptions.
Moreover, the broader long-term effects, such as the cumulative impact on content consumption patterns and industry outcomes, remain open for future investigation.

Despite these limitations, our study provides a strong foundation for further research into the role of LLM-generated explanations in recommender systems, emphasizing the importance of contextualization in improving user experience and engagement.

\section{Conclusion}
Our findings show how and why recommendation explanations meet cognitive, affective, and utilitarian motivations under the U\&G perspective. Our study confirms that LLM-generated contextualized explanations effectively address users' cognitive and affective needs and increase their consumption intentions. Notably, recommender systems should also avoid overwhelming users with irrelevant information that can hinder decision-making or erode trust. Future recommender system designs that integrate contextualized LLM-generated explanations can more effectively serve a wide range of user motivations, ultimately improving user satisfaction and leading to actionable engagement with the recommended content.

\bibliography{references}

\newpage
\appendix

\section{Recommendation Corpus Generation}
\label{appendix:corpus_generation}

\begin{table}[htbp]
\centering
\scriptsize
\begin{tabular}{ll}
\toprule
\multicolumn{2}{l}{\textbf{Recommendation Corpus Generation}} \\
\midrule
\multicolumn{2}{l}{\textbf{Input:} Movie shortlist $\mathcal{M}$} \\
\multicolumn{2}{l}{\textbf{Output:} Corpus $\mathcal{C}$ containing recommendations and explanations} \\
\midrule
1 & \textbf{for each} $m_i \in \mathcal{M}$ \textbf{do} \\
   & \quad $m_i$: Represents a single movie in the shortlist $\mathcal{M}$ \\
2 & \quad \textbf{for} $k = 1$ \textbf{to} 4 \textbf{do} \\
   & \quad $k$: The recommendation order (1st, 2nd, 3rd, or 4th) \\
3 & \qquad \textbf{if} $k=1$ \textbf{then} \\
4 & \qquad \quad Generate $r_{i,1}$ from $m_i$ \\
   & \qquad \quad $r_{i,k}$: The $k$-th recommendation for the $i$-th movie \\
5 & \qquad  \textbf{else} \\
6 & \qquad \quad Generate $r_{i,k}$ from $\{m_i, r_{i,1}, \dots, r_{i,(k-1)}\}$ \\
   & \qquad \quad Previous recommendations are incorporated to generate $r_{i,k}$ \\
7 & \qquad \textbf{end if} \\
8 & \qquad Generate $e_{i,k,c}$ and $e_{i,k,g}$ for $r_{i,k}$ \\
   & \qquad $e_{i,k,c}$: Contextualized Explanation for $r_{i,k}$ \\
   & \qquad $e_{i,k,g}$: Generic Explanation for $r_{i,k}$ \\
9 & \qquad Save $r_{i,k}$, $e_{i,k,c}$, and $e_{i,k,g}$ in $\mathcal{C}$ \\
   & \qquad Corpus $\mathcal{C}$ stores the recommendations and their explanations \\
10 & \quad \textbf{end for} \\
11 & \textbf{end for} \\
\bottomrule
\end{tabular}
\caption{Recommendation corpus preparation steps. The corpus $\mathcal{C}$ contains recommendations $r_{i,k}$ for each movie $m_i \in \mathcal{M}$, with contextualized explanations ($e_{i,k,c}$) and generic explanations ($e_{i,k,g}$) for each recommendation.}
\label{tab:corpus_generation}
\end{table}

Table~\ref{tab:corpus_generation} details the procedure we use to construct the recommendation corpus $\mathcal{C}$ from a given movie shortlist $\mathcal{M}$. For each movie $m_i \in \mathcal{M}$, the system generates a sequence of four recommendations, indexed by $k = 1$ to $4$.

At each step:
\begin{itemize}
    \item The variable $m_i$ denotes a single movie from the input shortlist.
    \item The index $k$ indicates the recommendation order (which, equivalently, is the number of movies in the users viewing history used to produce the recommendation).
    \item The system generates $r_{i,k}$, the $k$-th recommendation for movie $m_i$.
\end{itemize}

For the first recommendation ($k=1$), the system conditions only on the original movie $m_i$. For subsequent recommendations ($k>1$), it incorporates $m_i$ and all previously generated recommendations ($r_{i,1}, \dots, r_{i,(k-1)}$) to produce $r_{i,k}$. This iterative approach models a realistic, context-aware recommendation process.

After generating each recommendation $r_{i,k}$, the system produces two types of explanations for users in different conditions:
\begin{itemize}
    \item $e_{i,k,c}$: a \emph{contextualized explanation} that references the earlier items in the recommendation sequence;
    \item $e_{i,k,g}$: a \emph{generic explanation} that relates $r_{i,k}$ to the original movie $m_i$ without incorporating context.
\end{itemize}
Each recommendation $r_{i,k}$ and its corresponding explanations ($e_{i,k,c}$ and $e_{i,k,g}$) are added to the final corpus $\mathcal{C}$. This procedure ensures that the corpus captures both sequential recommendation dynamics and consistency across explanation styles.

\section{Interview Questions}
\label{appendix:interview}

\begin{itemize}
    \item \emph{``What made you trust or doubt the recommendation after seeing the explanation?''}

    \item \emph{``How did the explanation affect your overall feeling about the movie? Why did it make you feel that way?''}

    \item \emph{``How did the explanation assist (or fail to assist) you in understanding if the movie fit your interests or needs?''}

    \item \emph{``After reading the explanation, what influenced you most in deciding whether to watch the movie or not?''}

    \item \emph{``How do you think your profile, such as your demographics or viewing habits, influenced how you felt about the explanation you received?''}
    
\end{itemize}

\section{Supplementary Experiments}

\subsection{Clade 3.5 Sonnet}
\label{appendix:claude}
\begin{figure}[htbp]
  \centering
  \includegraphics[width=0.45\textwidth]{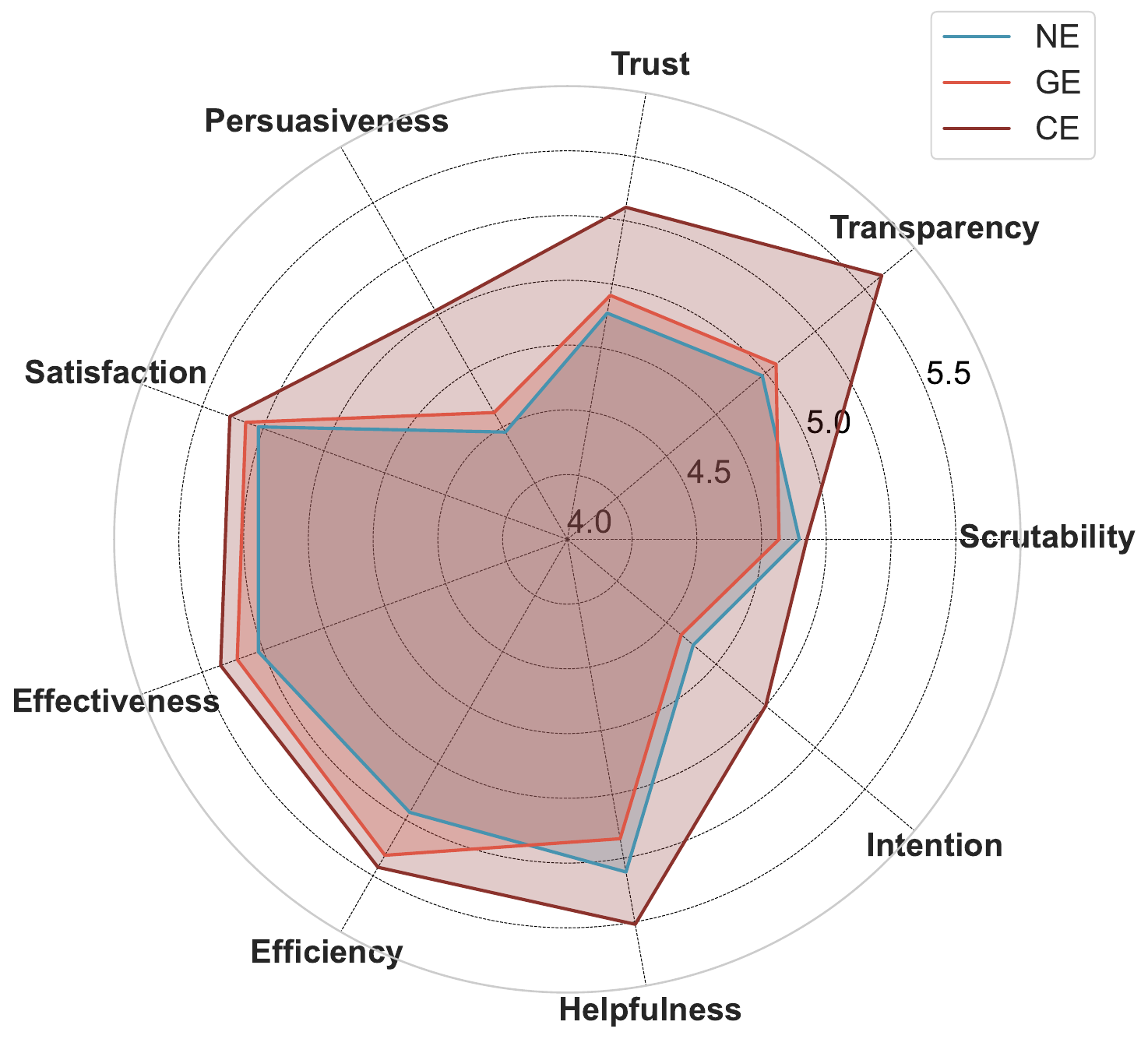}
  \caption{Radar chart of average user perception ratings in different dimensions by explanation types (\emph{Model: Claude 3.5 Sonnet}).}
  \label{fig:claude-radar}
\end{figure}

\begin{figure*}[h]
  \centering
  \includegraphics[width=1.0\textwidth]{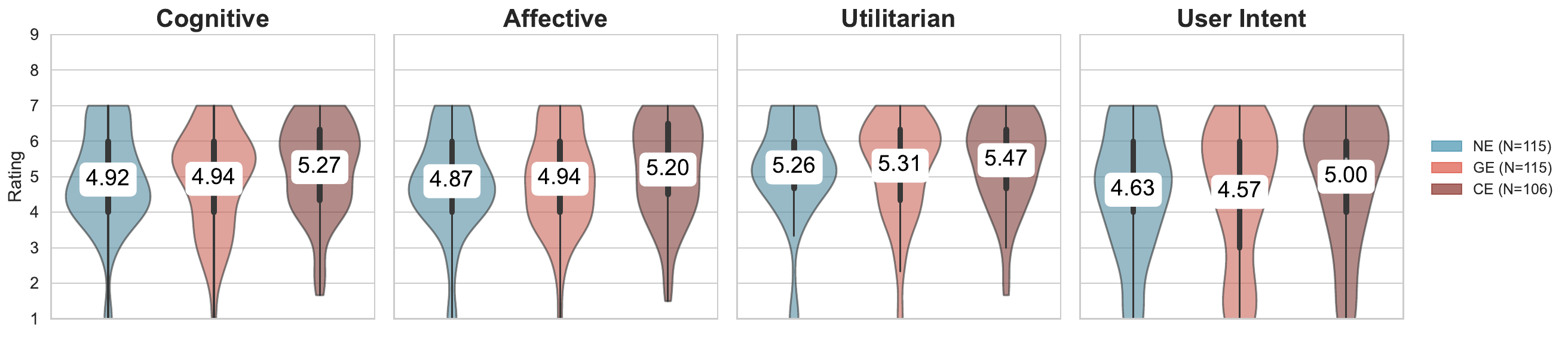}
  \caption{Ratings by explanation type (shown variable: \emph{Cognitive}, \emph{Affective}, \emph{Utilitarian}, and \emph{UserIntent}). Significance levels between conditions are indicated: $^*p<0.05$, $^{**}p<0.01$, $^{***}p<0.001$.}
  \label{fig:claude-violin}
\end{figure*}

We replicated the experiment on a smaller U.S. sample ($N$=140) using Claude 3.5 Sonnet. The results align with our main study, showing that contextualization improves overall explanation quality, particularly in terms of persuasiveness, transparency, and trust. Besides, contextualized explanations increase users' consumption intentions. While some minor discrepancies emerged between this supplementary experiment and the main experiment, the core findings remain consistent. Due to the limited sample size, some of the results are insignificant.
Overall, the results of the supplement experiment support the generalizability of our study's conclusions. Figure~\ref{fig:claude-violin} shows the user ratings of motivational needs and consumption intentions. Figure~\ref{fig:claude-radar} presents a more detailed comparison across dimensions. Table~\ref{table:mixed_effect_claude} shows the regression results.

\begin{table}[ht]
\centering
\scriptsize
\begin{tabular}{p{2cm}cccc}
\toprule
 & \emph{Cognitive} & \emph{Affective} & \emph{Utilitarian} & \emph{UserIntent} \\
\midrule
\emph{Intercept} &
\begin{tabular}[c]{@{}c@{}}4.955***\\ (0.198)\end{tabular} &
\begin{tabular}[c]{@{}c@{}}4.935***\\ (0.202)\end{tabular} &
\begin{tabular}[c]{@{}c@{}}5.280***\\ (0.204)\end{tabular} &
\begin{tabular}[c]{@{}c@{}}4.712***\\ (0.269)\end{tabular} \\

\emph{Treatment (GE)} &
\begin{tabular}[c]{@{}c@{}}$-$0.038\\ (0.249)\end{tabular} &
\begin{tabular}[c]{@{}c@{}}0.017\\ (0.254)\end{tabular} &
\begin{tabular}[c]{@{}c@{}}$-$0.200\\ (0.257)\end{tabular} &
\begin{tabular}[c]{@{}c@{}}$-$0.041\\ (0.340)\end{tabular} \\

\emph{Treatment (CE)} &
\begin{tabular}[c]{@{}c@{}}0.333\\ (0.255)\end{tabular} &
\begin{tabular}[c]{@{}c@{}}0.397\\ (0.260)\end{tabular} &
\begin{tabular}[c]{@{}c@{}}0.165\\ (0.263)\end{tabular} &
\begin{tabular}[c]{@{}c@{}}0.502\\ (0.348)\end{tabular} \\

\emph{Gender (Female)} &
\begin{tabular}[c]{@{}c@{}}$-$0.022\\ (0.209)\end{tabular} &
\begin{tabular}[c]{@{}c@{}}$-$0.190\\ (0.213)\end{tabular} &
\begin{tabular}[c]{@{}c@{}}0.105\\ (0.215)\end{tabular} &
\begin{tabular}[c]{@{}c@{}}$-$0.377\\ (0.285)\end{tabular} \\

\emph{Age} &
\begin{tabular}[c]{@{}c@{}}0.227*\\ (0.110)\end{tabular} &
\begin{tabular}[c]{@{}c@{}}0.196\\ (0.112)\end{tabular} &
\begin{tabular}[c]{@{}c@{}}0.239*\\ (0.113)\end{tabular} &
\begin{tabular}[c]{@{}c@{}}0.115\\ (0.150)\end{tabular} \\

\emph{Education} &
\begin{tabular}[c]{@{}c@{}}$-$0.043\\ (0.104)\end{tabular} &
\begin{tabular}[c]{@{}c@{}}$-$0.079\\ (0.106)\end{tabular} &
\begin{tabular}[c]{@{}c@{}}$-$0.090\\ (0.107)\end{tabular} &
\begin{tabular}[c]{@{}c@{}}0.001\\ (0.142)\end{tabular} \\

\emph{RecSysUtility} &
\begin{tabular}[c]{@{}c@{}}0.118\\ (0.156)\end{tabular} &
\begin{tabular}[c]{@{}c@{}}0.051\\ (0.159)\end{tabular} &
\begin{tabular}[c]{@{}c@{}}0.168\\ (0.160)\end{tabular} &
\begin{tabular}[c]{@{}c@{}}0.250\\ (0.214)\end{tabular} \\

\emph{RecSysFrequency} &
\begin{tabular}[c]{@{}c@{}}0.227\\ (0.185)\end{tabular} &
\begin{tabular}[c]{@{}c@{}}0.230\\ (0.188)\end{tabular} &
\begin{tabular}[c]{@{}c@{}}0.166\\ (0.190)\end{tabular} &
\begin{tabular}[c]{@{}c@{}}0.188\\ (0.252)\end{tabular} \\

\emph{MovieConsumption} &
\begin{tabular}[c]{@{}c@{}}0.099\\ (0.126)\end{tabular} &
\begin{tabular}[c]{@{}c@{}}0.044\\ (0.129)\end{tabular} &
\begin{tabular}[c]{@{}c@{}}$-$0.006\\ (0.130)\end{tabular} &
\begin{tabular}[c]{@{}c@{}}$-$0.092\\ (0.172)\end{tabular} \\

\emph{DisplayOrder} &
\begin{tabular}[c]{@{}c@{}}$-$0.077\\ (0.048)\end{tabular} &
\begin{tabular}[c]{@{}c@{}}$-$0.013\\ (0.050)\end{tabular} &
\begin{tabular}[c]{@{}c@{}}$-$0.011\\ (0.042)\end{tabular} &
\begin{tabular}[c]{@{}c@{}}$-$0.119\\ (0.082)\end{tabular} \\

\emph{HistoryNum} &
\begin{tabular}[c]{@{}c@{}}$-$0.001\\ (0.050)\end{tabular} &
\begin{tabular}[c]{@{}c@{}}$-$0.099\\ (0.052)\end{tabular} &
\begin{tabular}[c]{@{}c@{}}$-$0.003\\ (0.044)\end{tabular} &
\begin{tabular}[c]{@{}c@{}}$-$0.007\\ (0.084)\end{tabular} \\

\shortstack[l]{\emph{HistoryNum} \\ $\times$ \emph{MovieConsumption}} &
\begin{tabular}[c]{@{}c@{}}0.027\\ (0.055)\end{tabular} &
\begin{tabular}[c]{@{}c@{}}0.058\\ (0.057)\end{tabular} &
\begin{tabular}[c]{@{}c@{}}0.007\\ (0.048)\end{tabular} &
\begin{tabular}[c]{@{}c@{}}0.039\\ (0.091)\end{tabular} \\

\shortstack[l]{\emph{HistoryNum} \\ $\times$ \emph{RecSysUtility}} &
\begin{tabular}[c]{@{}c@{}}0.009\\ (0.057)\end{tabular} &
\begin{tabular}[c]{@{}c@{}}$-$0.040\\ (0.059)\end{tabular} &
\begin{tabular}[c]{@{}c@{}}$-$0.021\\ (0.050)\end{tabular} &
\begin{tabular}[c]{@{}c@{}}$-$0.006\\ (0.095)\end{tabular} \\
\bottomrule
\end{tabular}
\caption{Mixed-effects linear regression results. Coefficients are reported with standard errors in parentheses. Significance levels: $^*p<0.05$, $^{**}p<0.01$, $^{***}p<0.001$.}
\label{table:mixed_effect_claude}
\end{table}

\subsection{Movie Genre}
\label{appendix:mv_genre}
Observation of popular movie recommenders like Netflix and Amazon Video showed that most movies recommended tended to fall into widely favored genres such as \textit{Comedy} and \textit{Adventure}. Based on IMDb genre labels, we annotated each recommended movie in our experimental study as either \emph{Comedy} or \emph{Adventure}. To examine how genre influences user perceptions and consumption intentions, we performed the following regression analysis:
\begin{small}
\begin{align}
Y_{ij} &= \beta_0  
+ \beta_1 \emph{Treatment}_{j} 
+ \beta_2 \emph{Age}_{j} 
+ \beta_3 \emph{Gender}_{j} 
+ \beta_4 \emph{Education}_{j}\notag\\
&\quad + \beta_5 \emph{MovieConsumption}_{j} 
+ \beta_6 \emph{RecSysFrequency}_{j} \notag\\
&\quad + \beta_7 \emph{RecSysUtility}_{j} + \beta_8 \emph{HistoryNum}_{j} + \beta_9 \emph{DisplayOrder}_{j} \notag\\
&\quad + \beta_{10} \emph{MvGenre}_{ij} + \beta_{11} \emph{MvGenre}_{j} \times \emph{Treatment}_{j} 
+ u_{0j} + \epsilon_{ij} \notag
\end{align}
\label{eq:mixed_effect_mv_genre}
\end{small}

\noindent
where ${Y}_{ij}$ denotes the dependent variables (i.e., \emph{Cognitive}, \emph{Affective}, \emph{Utilitarian}, and \emph{UserIntent}) for the $i$-th observation of the $j$-th subject, with intercepts $\beta_0$, subject-level random effects $u_{0j}$, coefficients $\beta_1$ through $\beta_{11}$, and an error term $\epsilon_{ij}$.

\noindent
\textbf{Results} Regression results (Table~\ref{table:mixed_effect_genre}) show that the genre of the recommended movie significantly influenced user perceptions. Specifically, movies categorized as \emph{Comedy} were associated with lower ratings across affective ($\beta$=-0.203, $p$\textless0.05), utilitarian ($\beta$=-0.149, $p$\textless0.05), and behavioral intention outcomes ($\beta$=-0.415, $p$\textless0.05), compared to \emph{Adventure} movies. The effect of genre was not significantly moderated by explanation type (GE or CE), as the interaction terms were non-significant across all outcomes. These results suggest that genre perceptions persist regardless of contextualization strategies and that users may perceive adventure movies as more engaging or informative in recommendation explanations.

\begin{table}[ht]
\centering
\scriptsize
\begin{tabular}{p{2cm}cccc}
\toprule
 & \emph{Cognitive} & \emph{Affective} & \emph{Utilitarian} & \emph{UserIntent} \\
\midrule
\emph{Intercept} &
\begin{tabular}[c]{@{}c@{}}4.946***\\ (0.090)\end{tabular} &
\begin{tabular}[c]{@{}c@{}}5.045***\\ (0.090)\end{tabular} &
\begin{tabular}[c]{@{}c@{}}5.189***\\ (0.086)\end{tabular} &
\begin{tabular}[c]{@{}c@{}}4.879***\\ (0.117)\end{tabular} \\

\emph{Treatment (GE)} &
\begin{tabular}[c]{@{}c@{}}0.006\\ (0.106)\end{tabular} &
\begin{tabular}[c]{@{}c@{}}$-$0.058\\ (0.106)\end{tabular} &
\begin{tabular}[c]{@{}c@{}}0.034\\ (0.102)\end{tabular} &
\begin{tabular}[c]{@{}c@{}}$-$0.268\\ (0.138)\end{tabular} \\

\emph{Treatment (CE)} &
\begin{tabular}[c]{@{}c@{}}0.438***\\ (0.106)\end{tabular} &
\begin{tabular}[c]{@{}c@{}}0.135\\ (0.106)\end{tabular} &
\begin{tabular}[c]{@{}c@{}}0.125\\ (0.102)\end{tabular} &
\begin{tabular}[c]{@{}c@{}}0.188\\ (0.139)\end{tabular} \\

\emph{Gender (Female)} &
\begin{tabular}[c]{@{}c@{}}0.116\\ (0.088)\end{tabular} &
\begin{tabular}[c]{@{}c@{}}0.160\\ (0.086)\end{tabular} &
\begin{tabular}[c]{@{}c@{}}0.291***\\ (0.084)\end{tabular} &
\begin{tabular}[c]{@{}c@{}}$-$0.183\\ (0.110)\end{tabular} \\

\emph{Age} &
\begin{tabular}[c]{@{}c@{}}$-$0.075\\ (0.046)\end{tabular} &
\begin{tabular}[c]{@{}c@{}}$-$0.089*\\ (0.045)\end{tabular} &
\begin{tabular}[c]{@{}c@{}}0.013\\ (0.044)\end{tabular} &
\begin{tabular}[c]{@{}c@{}}$-$0.148*\\ (0.057)\end{tabular} \\

\emph{Education} &
\begin{tabular}[c]{@{}c@{}}0.037\\ (0.042)\end{tabular} &
\begin{tabular}[c]{@{}c@{}}0.028\\ (0.041)\end{tabular} &
\begin{tabular}[c]{@{}c@{}}0.002\\ (0.040)\end{tabular} &
\begin{tabular}[c]{@{}c@{}}0.038\\ (0.052)\end{tabular} \\

\emph{RecSysUtility} &
\begin{tabular}[c]{@{}c@{}}0.306***\\ (0.063)\end{tabular} &
\begin{tabular}[c]{@{}c@{}}0.298***\\ (0.062)\end{tabular} &
\begin{tabular}[c]{@{}c@{}}0.273***\\ (0.060)\end{tabular} &
\begin{tabular}[c]{@{}c@{}}0.297***\\ (0.079)\end{tabular} \\

\emph{RecSysFrequency} &
\begin{tabular}[c]{@{}c@{}}0.018\\ (0.072)\end{tabular} &
\begin{tabular}[c]{@{}c@{}}$-$0.004\\ (0.070)\end{tabular} &
\begin{tabular}[c]{@{}c@{}}$-$0.025\\ (0.069)\end{tabular} &
\begin{tabular}[c]{@{}c@{}}0.130\\ (0.089)\end{tabular} \\

\emph{MovieConsumption} &
\begin{tabular}[c]{@{}c@{}}0.152**\\ (0.047)\end{tabular} &
\begin{tabular}[c]{@{}c@{}}0.174***\\ (0.046)\end{tabular} &
\begin{tabular}[c]{@{}c@{}}0.174***\\ (0.045)\end{tabular} &
\begin{tabular}[c]{@{}c@{}}0.126*\\ (0.059)\end{tabular} \\

\emph{DisplayOrder} &
\begin{tabular}[c]{@{}c@{}}0.036\\ (0.024)\end{tabular} &
\begin{tabular}[c]{@{}c@{}}0.034\\ (0.028)\end{tabular} &
\begin{tabular}[c]{@{}c@{}}0.016\\ (0.025)\end{tabular} &
\begin{tabular}[c]{@{}c@{}}0.113**\\ (0.042)\end{tabular} \\

\emph{HistoryNum} &
\begin{tabular}[c]{@{}c@{}}0.045**\\ (0.015)\end{tabular} &
\begin{tabular}[c]{@{}c@{}}0.068***\\ (0.018)\end{tabular} &
\begin{tabular}[c]{@{}c@{}}0.042**\\ (0.015)\end{tabular} &
\begin{tabular}[c]{@{}c@{}}0.098***\\ (0.028)\end{tabular} \\

\emph{MvGenre (Com)} &
\begin{tabular}[c]{@{}c@{}}$-$0.096\\ (0.062)\end{tabular} &
\begin{tabular}[c]{@{}c@{}}$-$0.203**\\ (0.073)\end{tabular} &
\begin{tabular}[c]{@{}c@{}}$-$0.149*\\ (0.064)\end{tabular} &
\begin{tabular}[c]{@{}c@{}}$-$0.415***\\ (0.111)\end{tabular} \\

\shortstack[l]{\emph{MvGenre (Com)} \\ $\times$ \emph{Treatment (GE)}} &
\begin{tabular}[c]{@{}c@{}}0.014\\ (0.072)\end{tabular} &
\begin{tabular}[c]{@{}c@{}}0.064\\ (0.085)\end{tabular} &
\begin{tabular}[c]{@{}c@{}}0.077\\ (0.074)\end{tabular} &
\begin{tabular}[c]{@{}c@{}}0.096\\ (0.132)\end{tabular} \\

\shortstack[l]{\emph{MvGenre (Com)} \\ $\times$ \emph{Treatment (CE)}} &
\begin{tabular}[c]{@{}c@{}}$-$0.051\\ (0.071)\end{tabular} &
\begin{tabular}[c]{@{}c@{}}0.068\\ (0.084)\end{tabular} &
\begin{tabular}[c]{@{}c@{}}0.095\\ (0.073)\end{tabular} &
\begin{tabular}[c]{@{}c@{}}0.206\\ (0.131)\end{tabular} \\
\bottomrule
\end{tabular}
\caption{Mixed-effects linear regression results. Coefficients are reported with standard errors in parentheses. Significance levels: $^*p<0.05$, $^{**}p<0.01$, $^{***}p<0.001$.}
\label{table:mixed_effect_genre}
\end{table}

\end{document}